\documentclass[twocolumn,english,showpacs,prl]{revtex4}
\usepackage[T1]{fontenc}
\usepackage[latin9]{inputenc}
\usepackage{amsmath}
\usepackage{graphicx}
\usepackage{amssymb}
\usepackage{dcolumn}
\usepackage{bm}
\usepackage{amsfonts}
\usepackage{times}
\usepackage{babel}
\usepackage{babel}

\makeatletter
\@ifundefined{textcolor}{}
{
 \definecolor{BLACK}{gray}{0}
 \definecolor{WHITE}{gray}{1}
 \definecolor{RED}{rgb}{1,0,0}
 \definecolor{GREEN}{rgb}{0,1,0}
 \definecolor{BLUE}{rgb}{0,0,1}
 \definecolor{CYAN}{cmyk}{1,0,0,0}
 \definecolor{MAGENTA}{cmyk}{0,1,0,0}
 \definecolor{YELLOW}{cmyk}{0,0,1,0}
 }
\providecommand{\U}[1]{\protect\rule{.1in}{.1in}}
\makeatother
\makeatother

\begin{document}

\title{Model of $d$-wave electron pairing in hole doped cuprate
superconductors: A possible explanation for the pseudogap regime }
\author{X. Q. Huang$^{1,2}$}
\email{xqhuang@netra.nju.edu.cn}
\affiliation{$^{1}$Department of Physics and National Laboratory of Solid State
Microstructure, Nanjing University, Nanjing 210093, China\\
$^{2}$Department of Telecommunications Engineering ICE, PLAUST, Nanjing
210016, China }
\date{\today}

\begin{abstract}
The real-space localized hole pair is constructed in the Cu-O plane of the
hole-doped cuprate superconductors. We prove analytically and numerically
that two electrons, due to the nearest-neighbor Coulomb repulsive
confinement effect, can be in pairing inside a single plaquette with the $d$%
-wave symmetry. The scenario supports the `no glue' pairing picture for the
cuprates. Based on the scenarios, the physical origin of the Fermi pocket
(or Fermi arc) and the two pseudogap behavior are provided. Our framework
leads directly to a unified linear relationship between the pseudogap
temperature $T^{\ast }$ and the hole doping level $x$ in these compounds.
\end{abstract}

\maketitle

One of the most important open issues in high-$T_{c}$ superconductors is the
origin of the pseudogap and its relationship to superconductivity. It is
widely believed that the mysterious pseudogap \cite%
{Thomas1988,Alloul1989,Ding1996,Valla2006} may hold the key to understanding
the mechanism of the cuprate superconductors. In the past twenty years,
extensive experimental and theoretical efforts have been devoted to
understand the pseudogap phenomenon in the normal state of the underdoped
cuprates. Experimentally, angle-resolved photoemission spectroscopies
(ARPES), nuclear magnetic resonance (NMR) and scanning tunneling microscopy
(STM) have revealed many new and anomalous pseudogap-related physical
properties such as the $d$-wave pairing symmetry \cite{Ding1996}, the Fermi
pocket \cite{Marshall1996,Meng2009} or Fermi arc \cite{Norman1998,Meng2009},
the two pseudogap behavior \cite{Deutscher1999,Boyer2007} and the linear
dependent of the pseudogap temperature with hole doping level \cite%
{Gurvitch1987}. In addition, the experimental identification of the
pseudogaps in manganites \cite{Mannella2005} and the localized Cooper pairs
in insulating or nonmetallic materials \cite{Stewart2007} imply that there
is probably not any direct link between the pairing phenomena known as
pseudogaps and high-temperature superconductivity \cite{Pasupathy2008}.
Theoretically, even though a huge amount of research work on the pseudogap
has been conducted so far, yet its precise nature remains controversial.

Understanding the pairing mechanism of the pseudogap has been regarded as an
essential step towards elucidating the nature of the high-$T_{c}$
superconductivity. But in Anderson's view, the need for a bosonic glue to
pair electrons in cuprates is folklore rather than the result of scientific
logic \cite{Anderson2007}. If this is a correct notion, how can the strong
Coulomb repulsion between electrons be overcome to support the electron
pairing?

In this letter, we attempt to answer the above important question by
introducing a new model for $d$-wave pairing in hole-doped high-$T_{c}$
superconductors. It is shown analytically that the opening of the pseudogap
may originate merely from a real-space ultra short-range electron-ion
Coulomb interactions, and no any quasiparticle glues are needed in the
suggested pairing mechanism. The innovative approach offers an attractive
explanation for the $d$-wave pairing behavior as well as for other
pseudogap-related puzzles observed in hole-doped cuprates. This work
indicates that the complex physical phenomena may be well understood within
the most basic electromagnetic theory. Moreover, it is worth pointing out
that the proposed model not only provide new insights into the underlying
physics of the pseudogap, it could also lead to a breakthrough in the theory
of high-$T_{c}$ superconductivity.

\begin{figure}[hp]
\begin{centering}
\resizebox{0.95\columnwidth}{!}{ \includegraphics{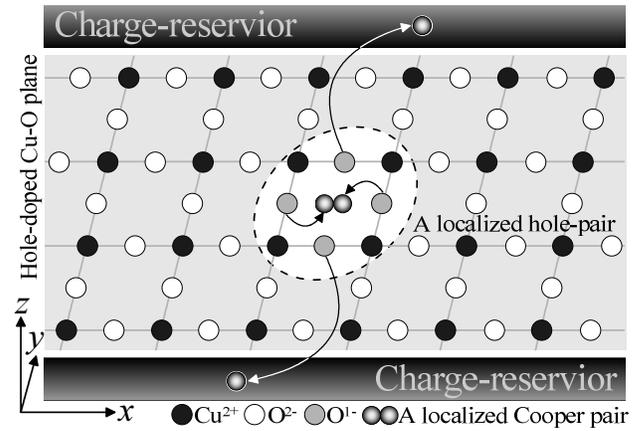}}
\par\end{centering}
\caption{The real-space structural relationship between the localized Cooper
pair and the localized hole pair.}
\label{fig1}
\end{figure}

In our opinion, a hole is a real-space `quasiparticle' which is composed of
some well-known electrons and ions. For the hole-doped cuprates, a localized
hole-pair is a cluster of two electrons (a localized Cooper pair), four $%
O^{1-}$ and four $Cu^{2+}$ inside the Cu-O plane, as illustrated in Fig. \ref%
{fig1}. In the following, we will show how this simply picture could yield a
pairing scenario that has the potential to resolve the pseudogap puzzle.

\begin{figure*}[tp]
\begin{centering}
\resizebox{1.3\columnwidth}{!}{ \includegraphics{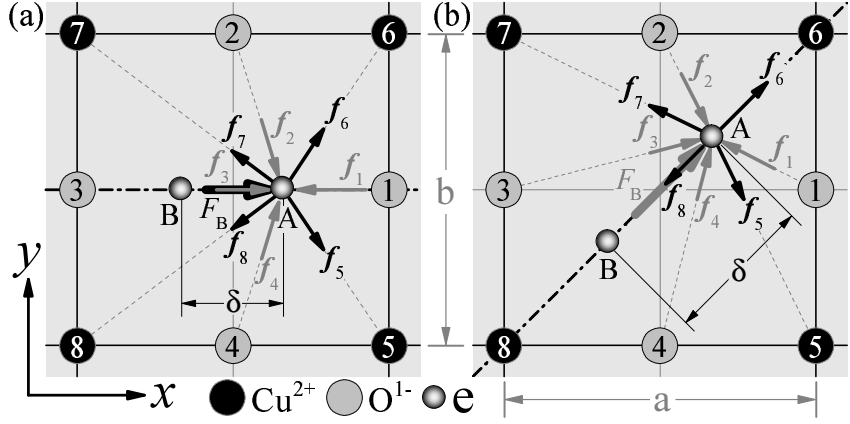}}
\par\end{centering}
\caption{The schematic plot of the confinement forces acting on the electron
pair (A and B) inside one unit cell of the Cu-O plane. Two special
situations are considered in this study, (a) two electrons arranged along
the $x$-direction, (b) two electrons aligned in the $xy$-direction. }
\label{fig2}
\end{figure*}

Figure \ref{fig1} immediately leads us to the question: how can two
repulsive electrons stay together inside a single plaquette? Here, we will
show that the real space nearest-neighbor electron-ion Coulomb interaction
is responsible for the mechanism (`pairing glue') of the pseudogap. As shown
in Fig. \ref{fig2}, two specific situations where two electrons (${A}$ and $%
B $) arranged on a line in $x$ and $xy$-direction are respectively
considered in the analysis. It can be seen from the figure, there are four
nearest-neighbor negative $O^{1-}$ ions (marked by 1, 2, 3, 4) and four
second nearest-neighbor positive $Cu^{2+}$ ions (marked by 5, 6, 7, 8)
around the electron pair. For the sake of simplicity we assume that $a=b$,
as a result, we can present the explicit analytical expressions of the
confinement forces on the electrons due to the structural symmetry. Based on
Fig. \ref{fig2}(a), the total confinement force $F_{x}$ applied to the
electron $A$ in $x$-direction takes the form
\begin{equation}
F_{x}=F_{B}+F_{x}^{(1)}+F_{x}^{(2)},  \label{force_y}
\end{equation}%
where the well-known Coulomb repulsion $F_{B}$ can be represented as $%
F_{B}=e^{2}/4\pi \varepsilon _{0}\delta ^{2}$. The parameter $\delta $ ($<b$%
) is the electron-electron spacing which can be used to characterize the
size of the Cooper pair. $F_{x}^{(1)}$ is the sum of forces caused by the
nearest-neighbor interactions (NNI) from four $O^{1-}$ ions and is given by
\begin{equation}
F_{x}^{(1)}=-\left\vert \mathbf{f_{1}}+\mathbf{f_{3}}\right\vert +\left\vert
\mathbf{f_{2}}+\mathbf{f_{4}}\right\vert =-\frac{e^{2}}{4\pi \varepsilon _{0}%
}\left( \frac{1}{d_{1}^{2}}-\frac{1}{d_{2}^{2}}\right) ,
\end{equation}%
with $d_{1}=(b^{2}-\delta ^{2})/4\sqrt{b\delta }$ and $d_{2}=\left(
b^{2}+\delta ^{2}\right) ^{3/4}/2\sqrt{2\delta }$. Similarly, the resultant
force $F_{x}^{(2)}$ originated from the second nearest neighbor interactions
(SNNI) of the four $Cu^{2+}$ ions can be expressed as%
\begin{equation}
F_{x}^{(2)}=\left\vert \mathbf{f_{5}}+\mathbf{f_{6}}\right\vert -\left\vert
\mathbf{f_{7}}+\mathbf{f_{8}}\right\vert =\frac{e^{2}}{2\pi \varepsilon _{0}}%
\left( \frac{1}{d_{3}^{2}}-\frac{1}{d_{4}^{2}}\right) ,
\end{equation}%
where the parameters $d_{3}$ and $d_{4}$ satisfy%
\begin{eqnarray}
d_{3} &=&\frac{(2b^{2}+\delta ^{2}-2b\delta )^{1/4}\sqrt{2b^{2}+\delta ^{2}-%
\sqrt{2}b\delta }}{4\sqrt{b-\delta }},  \notag \\
d_{4} &=&\frac{(2b^{2}+\delta ^{2}+2b\delta )^{1/4}\sqrt{2b^{2}+\delta ^{2}+%
\sqrt{2}b\delta }}{4\sqrt{b+\delta }}.  \label{d_14}
\end{eqnarray}%
While in the diagonal $xy$-direction, the total confinement force $F_{xy}$
has the following form
\begin{equation}
F_{xy}=F_{B}+F_{xy}^{(1)}+F_{xy}^{(2)},  \label{force_xy}
\end{equation}%
where $F_{B}=e^{2}/4\pi \varepsilon _{0}\delta ^{2}$ with $\delta <\sqrt{2}%
b. $ And the nearest neighbor resultant force $F_{xy}^{(1)}$ and the second
nearest neighbor resultant force $F_{xy}^{(2)}$ can be presented as
\begin{eqnarray}
F_{xy}^{(1)} &=&\frac{e^{2}}{4\pi \varepsilon _{0}}\left( \frac{1}{D_{1}^{2}}%
-\frac{1}{D_{2}^{2}}\right) ,  \label{f_xy1} \\
F_{xy}^{(2)} &=&\frac{e^{2}}{2\pi \varepsilon _{0}}\left( \frac{1}{D_{3}^{2}}%
-\frac{1}{D_{4}^{2}}\right) ,  \label{f_xy2}
\end{eqnarray}%
here the four distance parameters $D_{1}$, $D_{2}$, $D_{3}$ and $D_{4}$ are
given by
\begin{eqnarray}
D_{1} &=&\frac{(b^{2}+\delta ^{2}+\sqrt{2}b\delta )^{3/4}}{2\sqrt{\sqrt{2}%
b+2\delta }},\ D_{2}=\frac{(b^{2}+\delta ^{2}-\sqrt{2}b\delta )^{3/4}}{2%
\sqrt{\sqrt{2}b-2\delta }},  \notag \\
D_{3} &=&\frac{(2b^{2}-\delta ^{2})}{4\times 2^{3/4}\sqrt{b\delta }},\quad
D_{4}=\frac{\left( 2b^{2}+\delta ^{2}\right) ^{3/4}}{4\sqrt{\delta }}.
\label{D1_4}
\end{eqnarray}

\begin{figure}[tp]
\begin{centering}
\resizebox{0.9\columnwidth}{!}{ \includegraphics{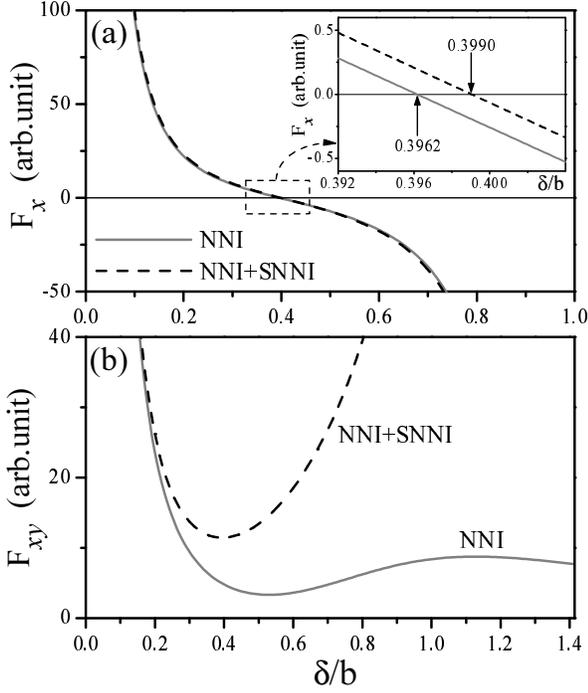}}
\par\end{centering}
\caption{Analytical confinement forces versus $\protect\delta /b$ in two
special directions (horizontal and diagonal), the gray solid lines represent
the results with the nearest neighbor interactions (NNI), while the black
dash lines are the numerical results of the nearest neighbor and second
nearest neighbor interactions (NNI+SNNI). (a) Two electrons in $x$%
-direction, as the case in Fig. \ref{fig2}(a), (b) two electrons in $xy$-direction, as the case in Fig. \ref{fig2}(b).}
\label{fig3}
\end{figure}

In the framework of Fig. \ref{fig2}, whether the two electrons become paired
inside the square lattice can be judged by the value of $F_{x}$ or $F_{xy}$.
If there exist a value of $\delta $ (electron-electron spacing) which can
ensure $F_{x}=0$ (or $F_{xy}=0$), then the pair can maintain it's integrity
in the single plaquette due to a complete elimination of the Coulomb
repulsion between electrons. With the analytical expressions from (\ref%
{force_y}) to (\ref{D1_4}), we calculate and draw in Fig. \ref{fig3} the
total forces ($F_{x}$ and $F_{xy}$) on the electron $A$ versus $\delta /b$
for the cases of NNI (the gray solid lines) and NNI+SNNI (the black dash
lines), respectively. As seen in Fig. \ref{fig3}(a), there exist always one $%
\delta $ with the force $F_{x}=0$ when the two electrons arranged along $x$%
-direction, moreover, the adding of the SNNI has little impact on the
formation of the stable electron pair in this direction (see the inset
figure). When two electrons arranged in $xy$-direction, the forces $F_{xy}$
are always positive for both NNI and NNI+SNNI, as shown in Fig. \ref{fig3}%
(b), these results imply that the electron-electron repulsion cannot be
entirely excluded if two electrons are aligned in $xy$-direction. Taking
into account the symmetry of Fig. \ref{fig2}, one can conclude from the
above discussions that two electrons may be glued together when they are
linked in both the horizontal ($x$ and -$x$) and vertical directions ($y$
and -$y$), surprisingly, the nearest-neighbor electron-ion repulsive
interactions can play the key role of the `pairing glue'. However, the two
electrons cannot form a bound state any longer when they stay in the four
diagonal directions ($xy$, $-xy$, $x$-$y$ and -$x$-$y$). These results imply
a possibility pseudogap phase of $d$-wave symmetry in the hole-doped
cuprates.

Except for the two special cases of Fig. \ref{fig2}, how about other
situations?

\begin{figure}[tp]
\begin{centering}
\resizebox{0.8\columnwidth}{!}{ \includegraphics{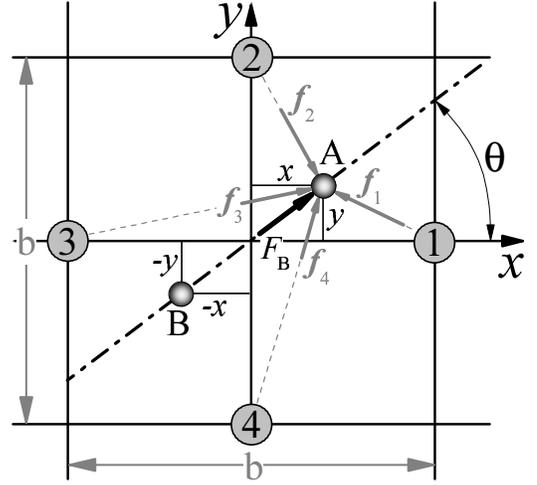}}
\par\end{centering}
\caption{The real space confinement effect on the two electrons (A and B)
located symmetrically at ($x$,$y$) and (-$x$,-$y$) inside the single
plaquette, where only the nearest-neighbor interactions are considered. }
\label{fig4}
\end{figure}

In Fig. \ref{fig3}, the numerical results show convincingly that the pairing
phenomenon is dominated by the nearest neighbor electron-ion interactions.
Hence, it is physically reasonable to consider only the NNI when one study
the real-space confinement effect in Cu-O plane, as shown in Fig. \ref{fig4}%
. Even though this physical picture appears simple, but it can reveal the
all the underlying physics of pseudogap behavior in cuprates. Classically,
the sum force on electron $A$ can be decomposed into its vertical and
horizontal components:
\begin{eqnarray}
F_{x} &=&\frac{e^{2}}{4\pi \varepsilon _{0}}\left( \frac{x-b/2}{\Delta
_{1}^{3}}+\frac{x}{\Delta _{2}^{3}}+\frac{x+b/2}{\Delta _{3}^{3}}+\frac{x}{%
\Delta _{4}^{3}}+\frac{x}{2\Delta _{B}^{3}}\right) ,  \notag \\
F_{y} &=&\frac{e^{2}}{4\pi \varepsilon _{0}}\left( \frac{y}{\Delta _{1}^{3}}+%
\frac{y-b/2}{\Delta _{2}^{3}}+\frac{y}{\Delta _{3}^{3}}+\frac{y+b/2}{\Delta
_{4}^{3}}+\frac{y}{2\Delta _{B}^{3}}\right) ,  \notag \\
&&  \label{FxFy}
\end{eqnarray}%
with $\Delta _{1}=\sqrt{y^{2}+(b/2-x)^{2}}$, $\Delta _{2}=\sqrt{%
x^{2}+(b/2-y)^{2}}$, $\Delta _{3}=\sqrt{y^{2}+(b/2+x)^{2}}$, $\Delta _{4}=%
\sqrt{x^{2}+(b/2+y)^{2}}$ and $\Delta _{B}=\sqrt{x^{2}+y^{2}}$ .

\begin{figure}[tp]
\begin{centering}
\resizebox{1\columnwidth}{!}{ \includegraphics{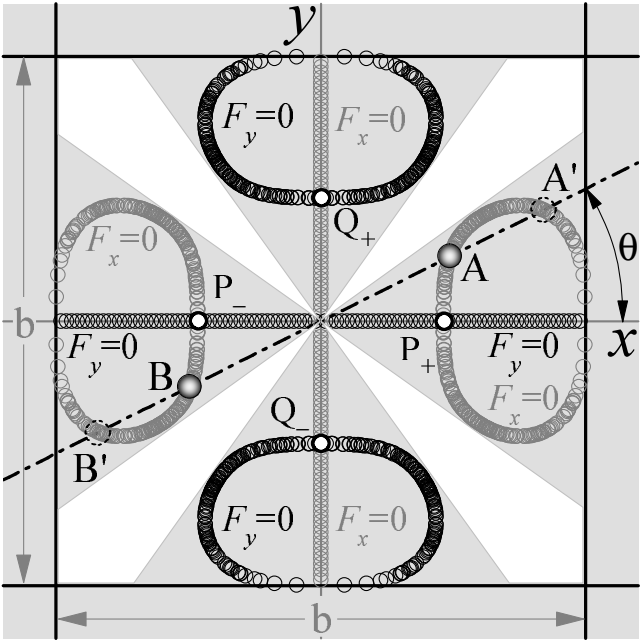}}
\par\end{centering}
\caption{The stability of the localized Cooper pair oriented in different
directions. The gray and black circles indicate the spatial locations at
which the resultant forces ($F_{x}$ and $F_{y}$) on the electrons in $x$-
and $y$-direction are zero, respectively. For more detail, see the text.}
\label{fig5}
\end{figure}

Of course, what concerns us most is whether the Coulomb repulsion between
electrons can be completely overcome in favor of the electron pairing.
Hence, we calculate with Eq. (\ref{FxFy}) and show in Fig. \ref{fig5} only
the results of $F_{x}=0$ (the gray circles) and $F_{y}=0$ (the black
circles). It is interesting to note that the locations of $F_{x}=0$ and $%
F_{y}=0$ form two real-space symmetrical `pocket-like' (or arc-like)
structures, respectively. In particular, there are two pair locations ($%
P_{+} $, $P_{-}$) and ($Q_{+}$, $Q_{-}$) where $F_{x}=F_{y}=0$ indicating a
complete elimination of the Coulomb repulsion inside the pairs and the
possibility of existence of the stable localized Cooper pairs, as already
confirmed in Fig. \ref{fig3}(a). However, the formation of the Cooper pairs
are forbidden around the diagonal directions (indicated by the white polygon
in the figure) because of $F_{x}\neq 0$ and $F_{y}\neq 0$. If two electrons (%
$A$ and $B$) locate symmetrically in the two `pockets' of same color (for
example, the gray pockets as indicated in the figure), the electron pair is
in a metastable state as the main repulsive component $F_{x}=0$, while the
minor component $F_{y}\neq 0$. Moreover, one can easily find from the figure
that, for a given $\theta $, there exist two metastable pseudogap states ($%
AB $ and $A^{\prime }B^{\prime }$) characterized by the pair's size $\delta
_{+}(\theta )=\overline{AB}$ and $\delta _{-}(\theta )=\overline{A^{\prime
}B^{\prime }}$, which may correspond to the two pseudogap behavior. Now we
can summarize qualitatively the main conclusions of Fig. \ref{fig5} into the
following expression of the pseudogap energy $E_{g}(\theta )$:

\begin{equation}
E_{g}(\theta )=\frac{1}{4\pi \varepsilon _{0}\delta _{\pm }(\theta )}%
\left\vert \cos ^{2}(\theta )-\sin ^{2}(\theta )\right\vert .  \label{Eg}
\end{equation}%
Physically, a large value of $E_{g}(\theta )$ usually corresponds a stable
localized Cooper pair which is experienced a strong confinement effect. It
is worth pointing out that Fig. \ref{fig5} and Eq. (\ref{Eg}) have already
included the three key experimental facts of the pseudogap phenomena: the $d$%
-wave pairing symmetry, Fermi pocket (or Fermi arc) and the two pseudogap
behavior. Furthermore, note that the physical properties of a system of two
Fermi electrons illustrated by Fig. \ref{fig5} are antisymmetric under the
exchange of the two identical electrons inside one pair.

Finally, we try to give a brief interpretation of the carrier doping ($x$)
dependence of the pseudogap temperature $T^{\ast }$ in the hole-doped
cuprates. Based on the simple scenario of Fig. \ref{fig1} and Fig. \ref{fig2}%
, in the zero-temperature approximation, an isolated single stable localized
Cooper pair can be defined by the binding energy $E_{b}^{s}(0)$ of the
pseudogap state. In a real system, the hypothesis of zero temperature and
single Cooper pair is physically untrue. Obviously, both temperature and the
Coulomb interaction between Cooper pairs would lead to a decreasing of the
stability of the localized Cooper pairs, which can be qualitatively
described by the following formula of the binding energy
\begin{equation}
E_{b}(T)=E_{b}^{s}(0)-\alpha k_{B}T-\frac{\beta }{\overline{\xi }_{ab}},
\label{Eb1}
\end{equation}%
where $\alpha $ and $\beta $ are constants, $k_{B}$ is Boltzmann's constant,
$T$ is the temperature in degrees Kelvin and  $\overline{\xi }_{ab}$ is the
average distance of the localized Cooper pairs inside one Cu-O plane. For
layered cuprates, the pair-pair interactions between different Cu-O layers
are negligible, hence, only the interactions inside one Cu-O plane are taken
into account in Eq. (\ref{Eb1}) through the parameter $\overline{\xi }_{ab}$%
. In the quasi-two-dimensional system, $\overline{\xi }_{ab}$ is roughly
inversely proportional to the doping level $x$: $\overline{\xi }%
_{ab}=\lambda /x$, where $\lambda $ is a constant.

Equation (\ref{Eb1}) means that there exists a critical temperature (or
pseudogap temperature) $T^{*}$ below which the binding energy $E_{b}(T)>0$,
the pseudogap may be expected in the superconductor, while above which $%
E_{b}(T)<0$, the pseudogap may disappear due to the complete destruction of
the localized Cooper pair. With the critical condition $E_{b}(T^{*})=0$,
then from Eq. (\ref{Eb1}) we obtain

\begin{equation}
T^{*}=\frac{1}{\alpha k_{B}}\left[E_{b}^{s}(0)-\frac{\beta}{\lambda}x\right].
\end{equation}

The above equation suggests that the pseudogap temperature $T^{*}$ decreases
linearly with an increase in doping level $x$, which is in good agreement
with the experiments.

In conclusion, we have proposed for the first time the model of the
real-space localized hole pair for the hole-doped cuprates. In the new
framework, the localized Cooper pair can naturally form inside a square
lattice of four $O^{1-}$ ions with the $d$-wave symmetry. The
nearest-neighbor characteristic of the real-space pairing mechanism
indicates that the pseudogap is a common natural phenomenon that can be
observed in various materials with low carrier concentration. The physical
origins of the Fermi pocket formation (or Fermi arc) and the two pseudogap
behavior have been uniquely determined by the scenarios developed. Finally,
the linear relationship between the pseudogap temperature $T^{\ast }$ and
the doping level $x$ in the hole-doped cuprates has been analytically
proved. The contents of these results are believed to be strong evidence for
the `no glue' picture which was first argued by Anderson \cite{Anderson2007}
and recently confirmed by experiment \cite{Pasupathy2008}. We are confident
that these findings have shed light on the unresolved problem of the
pseudogap.

The author acknowledge the valuable discussion with Professors P. W.
Anderson.


\begin{thebibliography}{99}
\bibitem{Thomas1988} G. A. Thomas \emph{et al}., Phys. Rev. Lett. \textbf{61}%
, 1313 (1988).

\bibitem{Alloul1989} J. W. Loram, K. A. Mirza, J. R. Cooper, and W. Y.
Liang, Phys. Rev. Lett. \textbf{71}, 1740 (1993).

\bibitem{Ding1996} H. Ding \emph{et al}., Nature \textbf{382}, 51 (1996).

\bibitem{Valla2006} T. Valla, A. V. Fedorov, Jinho Lee, J. C. Davis and G.
D. Gu, Science \textbf{314}, 1914 (2006).

\bibitem{Marshall1996} D. S. Marshall \emph{et al}., Phys. Rev. Lett. 76,
4841 (1996).

\bibitem{Meng2009} J. Q. Meng \emph{et al}., Nature \textbf{462}, 335 (2009).

\bibitem{Norman1998} M. Norman \emph{et al}., Nature \textbf{392}, 157
(1998).

\bibitem{Deutscher1999} G. Deutscher, \emph{et al}., Nature \textbf{397},
410 (1999).

\bibitem{Boyer2007} M. C. Boyer, \emph{et al}., Nature Phys. \textbf{3}, 802
(2007).

\bibitem{Gurvitch1987} M. Gurvitch and A. T. Fiory, Phys. Rev. Lett. \textbf{%
59}, 1337 (1987).

\bibitem{Mannella2005} N. Mannella \emph{et al}., Nature \textbf{438}, 474
(2005).

\bibitem{Stewart2007} M. D. Stewart Jr., Aijun Yin, J. M. Xu, James M.
Valles Jr., Science \textbf{318}, 1273 (2007).

\bibitem{Pasupathy2008} A. N. Pasupathy \emph{et al}., Science \textbf{320}
196 (2008).

\bibitem{Anderson2007} P. W. Anderson, Science \textbf{317}, 1705 (2007).
\end{thebibliography}
\end{document}